\documentclass{ws-ijmpe}

\begin{document}

%\markboth{Authors' Names}{Instructions for  
%Typing Manuscripts (Paper's Title)}

%%%%%%%%%%%%%%%%%%%%% Publisher's Area please ignore %%%%%%%%%%%%%%%
%\catchline{}{}{}{}{}
%%%%%%%%%%%%%%%%%%%%%%%%%%%%%%%%%%%%%%%%%%%%%%%%%%%%%%%%%%%%%%%%%%%%

\title{EXPLANATION OF THE RHIC HBT PUZZLE BY A GRANULAR\\ SOURCE OF
QUARK-GLUON PLASMA DROPLETS}

\author{\footnotesize WEI-NING ZHANG}

\address{Physics Department, Dalian University of Technology, Dalian, 
Liaoning 116024, China\\
Physics Department, Harbin Institute of Technology, Harbin, 
Heilongjiang 150006, China\\
weiningzh@hotmail.com}

\author{CHEUK-YIN WONG}

\address{Physics Division, Oak Ridge National Laboratory, 
Oak Ridge, TN 37831, USA\\
Department of Physics, University of Tennessee, Knoxville, 
TN 37996, USA\\
wongc@ornl.gov}

\maketitle

\begin{history}
\received{(received date)}
\revised{(revised date)}
%\accepted{(Day Month Year)}
%\comby{(xxxxxxxxxx)}
\end{history}

\begin{abstract}
  We present a review on the explanation of the RHIC HBT puzzle by a
  granular pion-emitting source of quark-gluon plasma droplets.  The
  evolution of the droplet is described by relativistic hydrodynamics
  with an equation of state suggested by lattice gauge results.  The
  granular source evolution is obtained by superposing all of the
  evolutions of individual droplets.  Pions are assumed to be emitted
  thermally from the droplets at the freeze-out configuration
  characterized by a freeze-out temperature $T_f$.  We find that the
  average particle emission time scales with the initial radius of the
  droplet.  Pions will be emitted earlier if the droplet radius is
  smaller.  An earlier emission time will lead to a smaller extracted
  HBT radius $R_{\rm out}$, while the extracted HBT radius $R_{\rm
  side}$ is determined by the scale of the distribution of the droplet
  centers.  However, a collective expansion of the droplets can
  further decrease $R_{\rm out}$.  As a result, the value of $R_{\rm
  out}/R_{\rm side}$ can be close to, or even less than 1 for the
  granular source of QGP droplets.
\end{abstract}

\section{Introduction}

HBT (Hanbury-Brown-Twiss) interferometry is an indispensable tool to
study the space-time structure of the particle-emitting source
produced in high energy heavy ion collisions\cite{Won94,Wie99,Wei00}.
The experimental pion HBT measurements at RHIC give the ratio of
$R_{\rm out}/ R_{\rm side} \approx
1\,$\cite{STA01a,PHE02a,PHE04a,STA05a}, which is much smaller than
many earlier theoretical expectations. Such a discrepancy between
theory and experiment is referred to as the RHIC HBT
puzzle\cite{Wie99,Wei00,Ris96,Tea99,Sof01,Pra03}.  On the other hand,
hydrodynamical calculations give reasonably good descriptions of the
elliptic flow, which has been considered as an evidence for a
strongly-coupled quark-gluon
plasma\cite{Gyu05,BRA05,PHO05,STA05,PHE05}.  The resolution of the HBT
puzzle is important in finding out why traditional hydrodynamics
succeed in explaining the elliptic flow but fails in explaining the
HBT radii.

Traditional studies of the hydrodynamics of the evolving fluid assume
a single contiguous blob of matter under expansion, with a relatively
smooth initial and final density distributions.  Initial transverse
density fluctuations and hydrodynamical instabilities have been
neglected but their inclusion may lead to ``multi-fragmentation'' in
the form of large scale final-state density fluctuations and the
formation of granular droplets.  It is useful to explore the
consequences of the occurrence of granular droplets.

Previously we propose a granular model to explain the HBT
puzzle\cite{Zha04,Zha06}. We would like to review here the important
ingredients which enters into the resolution of the puzzle.  Further
suggestions of using single-event HBT interferometry to search for
signatures of the granular source can be found in Refs.\
[\cite{Won04,Zha05,Won06}].

\section{Evolution of a QGP droplet}

Based on the recent results of high-energy heavy-ion collisions at
RHIC, the early matter produced in the collisions may be a
strongly-coupled QGP (sQGP), which has a very high energy density and
reaches local thermalization within about 1
fm/c\cite{Gyu05,BRA05,PHO05,STA05,PHE05}.  The expansion of the matter
after that time may be unstable.  Many effects, such as the large
fluctuations of the initial transverse energy
density\cite{Dre02,Soc04,Ham05}, the sausage instability\cite{Won73},
and possible phase transition\cite{Wit84}, may lead to the
fragmentation of the system and the formation of many spherical
droplets due to the surface tension of the QGP \cite{Zha06,Won04,Won06}.

To describe the evolution of a droplet, we use relativistic
hydrodynamics where the energy momentum tensor of a thermalized fluid
element in the center-of-mass frame of the droplet is\cite{Lan59}
\begin{equation}
\label{tensor}
T^{\mu \nu} (x') = \big [ \epsilon(x') + p(x') \big ] u^{\mu}(x') 
u^{\nu}(x') - p(x') g^{\mu \nu} \, ,
\end{equation}
$x'$ is the space-time coordinate of the fluid element in the
center-of-mass frame, $\epsilon$, $p$, and $u^{\mu}=\gamma
(1,{\hbox{\boldmath $v$}})$ are the energy density, pressure, and
4-velocity of the element, and $g^{\mu \nu}$ is the metric tensor.
With the local conservation of energy and momentum, one can obtain the
equations for spherical geometry as\cite{Ris96}
\begin{eqnarray}
\label{eqe}
\partial_t E + \partial_r [(E+p)v]  = - F \, , 
\end{eqnarray}
\begin{eqnarray}
\label{eqm}
\partial_t M + \partial_r (Mv+p)  = - G \, ,
\end{eqnarray}
where $E \equiv T^{00}$, $M \equiv T^{0r}$, $F=2v(E+p)/r$, $G=2vM/r$. 

In the equations of motion (\ref{eqe}) and (\ref{eqm}) there are three
unknown functions $\epsilon$, $p$, $v$.  In order to obtain the
solution of the equations of motion, we need an equation of state
which gives a relation $p(\epsilon)$ between $p$ and $\epsilon$
[\cite{Ris96,Ris98}].  At RHIC energy, the system undergoes a
transition from the QGP phase to hadronic phase.  As the net baryon
density in the central rapidity region is much smaller than the energy
density of the produced matter (here presumed to be QGP), the baryon
density of the system in the center rapidity region can be neglected.
Lattice gauge results suggest the entropy density of the system as a
function of temperature as\cite{Ris96,Ris98,Bla87,Lae96}
\begin{equation}
\label{eos}
{s(T) \over s_c} = \Bigg({T\over {T_c}}\Bigg)^{\!\!3} \Bigg[1+{{d_Q\!-\!d_H}
\over {d_Q\!+\!d_H}}\tanh\!\Bigg({{T\!-\!T_c}\over{\Delta T}}\Bigg)\Bigg]\,,
\end{equation}
where $s_c$ is the entropy density at the transition temperature $T_c$, 
$d_Q$ and $d_H$ are the degrees of freedom in the QGP phase and the hadronic 
phase, and $\Delta T$ is the width of the transition.  The thermodynamical 
relations among $p$, $\epsilon$, and $s$ in this case are
\begin{equation}
-sdT + dp=0\,,~~~~~~~~\epsilon = Ts-p \,.
\end{equation}
From these thermodynamical relations and Eq. (\ref{eos}), we can obtain
the equation of state $p(\epsilon)$.

\vspace*{-0.6cm}
\begin{figure}[h]
\hspace*{2.8in}
\psfig{file=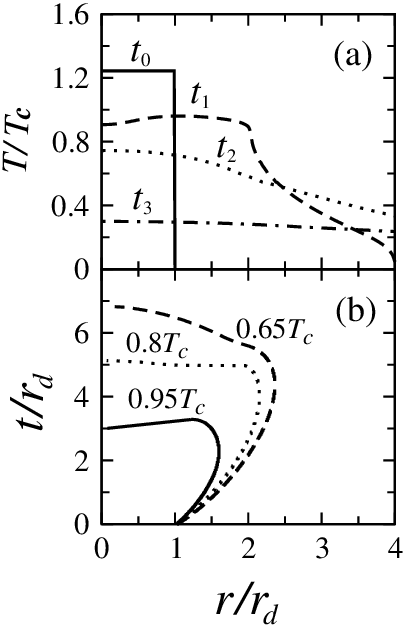,width=5cm}
\noindent 
\end{figure}

\vspace*{-0.7cm}
\hangafter=0
\hangindent=2.85in
\noindent Fig. 1.  
(a) Temperature profile and (b) isotherms for the droplet. 
Here, $t_n = 3n\lambda r_d$ and $\lambda=0.99$.

\vspace*{-8.8cm} \hangafter=-16 \hangindent=-2.3in Using the HLLE
scheme\cite{Sch93,Ris95} and Sod's operator splitting
method\cite{Sod77}, one can obtain the solution of Eqs. (\ref{eqe})
and (\ref{eqm})\cite{Ris96,Ris98,Zha04}, after knowing the equation of
state and initial conditions.  We assume that the droplet has a
uniform initial energy density $\epsilon_0$ within a sphere with
radius $r_d$, and has a zero initial velocity in its center-of-mass
frame.  Figs.\ 1(a) and (b) show the temperature profiles and
isotherms for the droplet.  In our calculations, we take the
parameters of the equation of state as $d_Q=37$, $d_H=3$, $T_{c}=165$
MeV, and $\Delta T = 0.05\ T_{c}$, and take the initial energy density
$\epsilon_0 =3.75 T_c s_c$, which is about two times of the density of
quark matter at $T_c$ [\cite{Ris96,Ris98}].

\section{Granular Source of QGP droplets}

\hangafter=-3 \hangindent=-2.3in If we assume that the final pions are
emitted from the droplet at the freeze-out configuration characterized
by a freeze-out temperature $T_f$, we can see from figure 1(b) that
the the average particle emission time scales with the initial radius
of the droplet $r_d$.  In HBT interferometry, the radius $R_{\rm
  side}$ is related to the spatial size of the particle-emitting
source and the radius $R_{\rm out}$ is related not only to the source
spatial size but also to the lifetime of the
source\cite{Wie99,Wei00,Pra90,Ber88}.  A long lifetime of the source
will lead to a large $R_{\rm out}$\cite{Wie99,Wei00,Pra90,Ber88}.
From the hydrodynamical solution in figure 1(b), both the average
freeze-out time and freeze-out radial distance increase with $r_d$ for
a single droplet source.  As a consequence, $R_{\rm out} / R_{\rm
  side}$ is insensitive\cite{Zha04} to the values $r_d$.  The value of
$R_{\rm out} / R_{\rm side}$ for the single droplet source\cite{Zha04}
is about 3 [\cite{Zha04}], much larger than the observed
values\cite{STA01a,PHE02a,PHE04a,STA05a}.

The RHIC HBT puzzle of $R_{\rm out} / R_{\rm side}\sim 1$ suggests
that the pion emitting time may be very short.  In order to explain
the HBT puzzle, we consider a granular source of $N_d$ QGP droplets
distributed in a distribution $D(X_d)$ as illustrated in Fig.\ 2(a).
We assume that the evolution of the granular source is simply the
superposition of the evolutions of individual droplets with the same
initial conditions.  As the average freeze-out time is proportional to
the initial radius of the droplet, the freeze-out time and $R_{\rm
  out}$ decreases if the initial radius of the droplet decreases.  On
the other hand, $R_{\rm side}$ increases if the width of the droplet
spatial distribution $D(X_d)$ increases.  A variation of the droplet
size and the width of droplet spatial distribution can result in
$R_{\rm out}$ nearly equal to $R_{\rm side}$ [\cite{Zha04}].
Furthermore, if the granular source has a collective expansion the HBT
will measure the size of the region of the ellipse in figure
2(b)\cite{Wie99,Wei00}.  In this case the value of $R_{\rm out}/R_{\rm
  side}$ decreases even further and may be smaller than
unity\cite{Zha04}.

\setcounter{figure}{1}       

\vspace*{-0.6cm}
\begin{figure}[th]
\centerline{\psfig{file=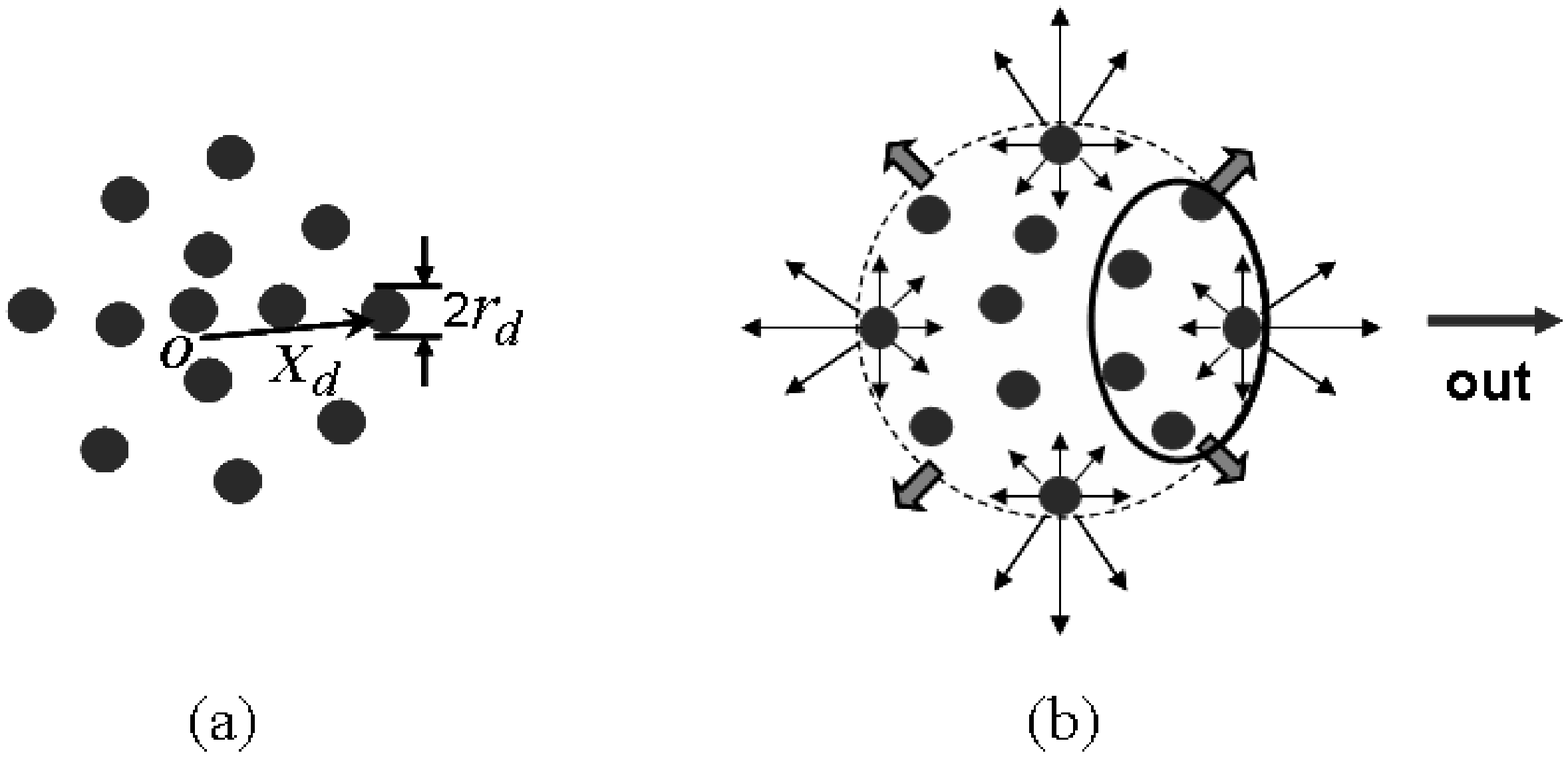,width=11cm}}
\vspace*{-0.2cm}
\caption{(a) Static Granular source of the QGP droplets.  (b) Expanding 
granular source of the QGP droplets.}
\end{figure}

\vspace*{-0.2cm} We assume that the droplets are initially distributed
 in a short cylinder of length $2{\cal R}_z$ along the beam direction
 ($z$ direction) with an initial transverse spatial distribution up to
 a radius ${\cal R}_t$,
\begin{eqnarray}
\frac{dP_d}{2\pi \rho \,d\rho \,dz} \propto [1-\exp(-\rho^2 /
\Delta{\cal R}_t^2)] \theta({\cal R}_t -\rho) \,\theta({\cal R}_z -|z|) \,,
\end{eqnarray}
where ${\bf \rho}=\sqrt{x^2+y^2}$ and $z$ are the coordinates of the
center of a droplet, and $\Delta {\cal R}_t$ describes a shell-type
radial distribution.  Because of the early thermalization and the
anisotropic pressure gradient, the droplets will acquire anisotropic
initial velocities.  We therefore consider a granular source of the
the QGP droplets with an anisotropic velocity distribution and the
velocity of a droplet depends on the initial coordinates of the
droplet center, $(r_1, r_2, r_3)=(x,y,z)$, in the form
\begin{eqnarray}
\label{bet}
{\hbox{\boldmath $\beta$}}_i = a_i \,{\rm sign}(r_i) \bigg(\frac{|r_i|}
{{\cal R}_i} \bigg )^{\!\!b_i} \,, ~~~~~~~~~i=1,2,{\rm ~and~}3,
\end{eqnarray}
where $a_i$ describes the magnitude of the anisotropic expansion,
${\rm sign}(r_i)$ denotes the sign of $r_i$, and $b_i\,(b_{_T},\,b_z)$
are the exponential power parameters that describe the variation of
the velocities with $r_i$.  In our calculations the velocity
parameters are taken as $a_x=0.415$, $a_y=0.315$, $a_z=0.850$,
$b_x=b_y=0.42$, and $b_z=0.03$, by comparing with the experimental
data of pion transverse momentum spectral\cite{PHE04b} and elliptic
flow\cite{PHE03} in $\sqrt{s_{NN}}=200$ GeV Au + Au collisions at
RHIC.  We take the freeze-out temperature as $T_f=0.95T_c$ in our
calculations.

\section{HBT Results of the Granular Source of QGP Droplets}

The two-particle Bose-Einstein correlation function is defined as the
ratio of the two-particle momentum distribution $P(p_1,p_2)$ relative
to the the product of the single-particle momentum distribution
$P(p_1) P(p_2)$.  For a chaotic pion-emitting source, $P(p_i)~(i=1,2)$,
and $P(p_1,p_2)$ can be expressed as\cite{Won94}
\begin{eqnarray}
P(p_i) = \sum_{X_i} A^2(p_i,X_i) \,,~~~~~~
P(p_1,p_2) = \sum_{X_1, X_2} \Big|\Phi(p_1, p_2; X_1,
X_2 )\Big|^2 ,
\end{eqnarray}
where $A(p_i,X_i)$ is the magnitude of the amplitude for emitting a
pion with 4-momentum $p_i=({\hbox{\boldmath $p$}}_i,E_i)$ in the 
laboratory frame at $X_i$ and is given by the Bose-Einstein distribution 
with freeze-out temperature $T_f$ in the local rest frame of the source 
point.  $\Phi(p_1, p_2; X_1, X_2 )$ is the two-pion wave function.  
Neglecting the absorption of the emitted pions by other droplets, 
$\Phi(p_1, p_2;X_1, X_2 )$ is simply
\begin{eqnarray}
\label{PHI}
\Phi(p_1, p_2; X_1, X_2 ) =\frac{1}{\sqrt{2}} \Big[ A(p_1, X_1)
A(p_2, X_2) e^{i p_1 \cdot X_1 + i p_2 \cdot X_2} 
+ (X_1\leftrightarrow X_2) \Big]\, .
\end{eqnarray}
Using the components of ``out", ``side", and ``long"\cite{Pra90,Ber88}
of the relative momentum of the two pions, $q=|{\bf p_1}-{\bf p_2}|$,
as variables, we can construct the correlation function $C(q_{\rm
  out}, q_{\rm side},q_{\rm long})$ from $P(p_1,p_2)$ and
$P(p_1)P(p_2)$ by picking pion pairs from the granular source and
summing over ${\bf p_1}$ and ${\bf p}_2$ for each $(q_{\rm out},
q_{\rm side},q_{\rm long})$ bin\cite{Zha04}.  The HBT radii $R_{\rm
  out}$, $R_{\rm side}$, and $R_{\rm long}$ can then be extracted by
fitting the calculated correlation function $C(q_{\rm out},q_{\rm
  side},q_{\rm long})$ with the following parametrized correlation
function
\begin{equation}
\label{cq}
C(q_{\rm out},q_{\rm side},q_{\rm long})=1+\lambda \, e^{-q_{\rm out}^2
R_{\rm out}^2 -q_{\rm side}^2 R_{\rm side}^2 -q_{\rm long}^2 R_{\rm long}^2}
\,.
\end{equation}
                                                                                
\begin{figure}[th]
\centerline{\psfig{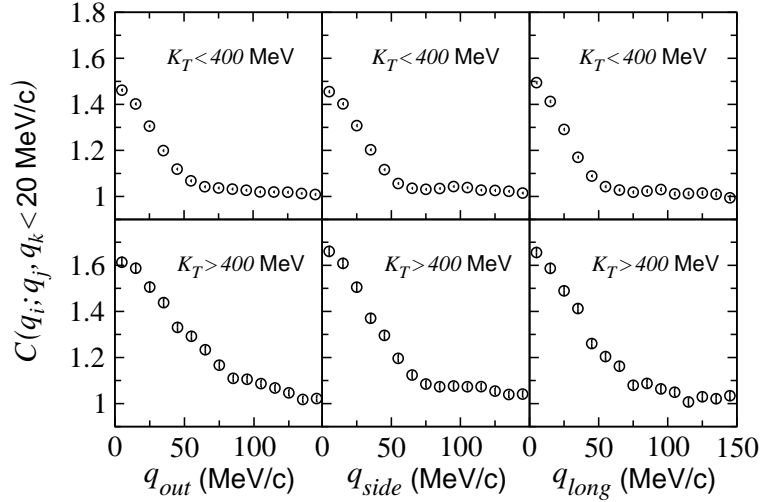}}
\vspace*{8pt}
\caption{Two-pion correlation functions for granular source.}
\end{figure}

\begin{figure}[th]
\centerline{\psfig{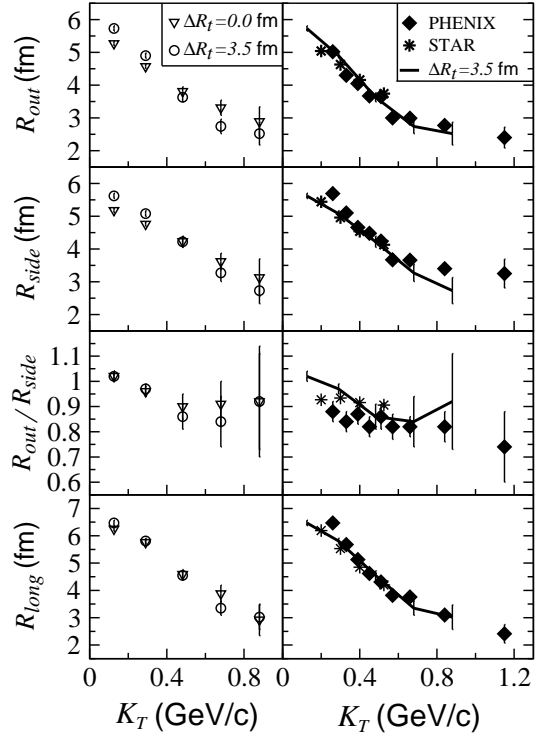}}
\vspace*{8pt}
\caption{Two-pion HBT radii obtained by the PHENIX 
Collaboration$^6$ and the STAR Collaboration$^7$
compared with the theoretical results calculated in the granular droplet 
model with $\Delta{\cal R}_t=0$ and 3.5 fm.}
\end{figure}

Figure 3 shows the theoretical two-pion correlation functions for the
granular source.  The top figures give the results for the average pion
transverse momentum of a pion pair, $K_T$, less than 400 MeV/c, and
the bottom figures for $K_T > 400$ MeV/c.  In our calculations the
size parameters of the granular sources are taken to be ${\cal
  R}_t=8.8$ fm, $\Delta{\cal R}_t=3.5$ fm, ${\cal R}_z=7.0$ fm, and
$r_d=1.3$ fm.  The number of droplet $N_d$ is taken to be 40 in our
calculations.  The left panels in Fig.\ 4 give the extracted two-pion
HBT radii for the granular source as a function of $K_T$.  The symbols
of circle and down-triangle are for $\Delta{\cal R}_t =0.35$ fm and
$\Delta{\cal R}_t =0$ fm, respectively.  The experimental PHENIX
results\cite{PHE04a}, and STAR results\cite{STA05a} are shown on the
right panels.  The curve gives the theoretical results for
$\Delta{\cal R}_t =0.35$ fm.  For our theoretical HBT calculations, we
use a cut for particle pseudo-rapidity region $|\eta|< 0.35$, the same
as in the PHENIX experiments\cite{PHE04a}.  We find that if we
increase the parameter $r_d$, the HBT radii $R_{\rm out}$ and $R_{\rm
  long}$ will increase.  And if we increase the parameter $\Delta{\cal
  R}_t$, the variation of HBT radii $R_{\rm out}$ and $R_{\rm side}$
with $K_T$ will become steep.  As can be inferred from Fig.\ 4, the
HBT results of granular source for $\Delta{\cal R}_t=3.5$ fm agree
quite well with experimental data, although the case with $\Delta
{\cal R}_t=0$ also give almost as good an agreement.
                                                                                
\section{Conclusions and Discussion}
                                                                                
The expansion of the dense matter (sQGP) produced in high-energy
heavy-ion collisions at RHIC may be unstable.  The initial transverse
density may also be highly fluctuation.  The unstable expansion and
fluctuating initial transverse density may lead to a fragmentation of
the system and the formation of a granular source of QGP droplets.
Although a granular structure was suggested earlier as the signature
of first-order phase transitions\cite{Wit84}, the occurrence of
granular structure may not be limited to the occurrence of first-order
phase transitions. There are additional effects which may lead to the
dynamical formation of granular droplets\cite{Zha06,Won06}.

For a granular source of the droplets, the average particle emission
time scales with the initial radius of the droplet.  Pions will be
emitted earlier if the radius is smaller.  An earlier emission time
will lead to a smaller extracted HBT radius $R_{\rm out}$, while the
extracted HBT radius $R_{\rm side}$ is determined by the scale of the
distribution of the droplets.  A collective expansion of the droplets
can further decrease $R_{\rm out}$ and the value of $R_{\rm
out}/R_{\rm side}$ can be close to, or even less than 1 for an
expanding granular source of QGP droplets.

In an event-mixing experimental analysis, some signals of the granular
source such as the correlation function fluctuations may likely be
suppressed after averaging over many events\cite{Won04}.  However, the
property of an earlier average emission time of the granular source
will remain as it is common to all events and not averaged out by
event mixing.  This property leads to a smaller $R_{\rm out}/R_{\rm
  side}$ in mixed-event HBT analysis.

In conclusion, a granular source model can explain the RHIC HBT puzzle
and reproduce the data of pion HBT radii, as well as the data of pion
transverse momentum spectra and elliptic flow\cite{Zha06} in
$\sqrt{s_{NN}}=200$ GeV Au + Au collisions at RHIC.  It is of great
interest to find direct evidences of the granular structure and to
study the mechanics of granular source
formation\cite{Zha04,Zha06,Won04,Zha05,Won06}.

\section*{Acknowledgments}

This research was supported by the National Natural Science Foundation of 
China under Contracts No. 10575024, and by the Division of Nuclear Physics, 
US DOE, under Contract No. DE-AC05-00OR22725 managed by UT-Battle, LC.

\end{document}